\documentclass[a4wide,vsecnum]{article}
\usepackage{longtable}

\input{tcilatex}

\begin{document}

\title{}
\author{\ F.F. Goryayev, A.M. Urnov and L.A. Vainshtein \\
%EndAName
\textit{Lebedev Physical Institute of RAS, Moscow 119991, Russia}}
\maketitle
\title{\textbf{Atomic data calculations by Z-expansion method for doubly
excited states }$\mathbf{2lnl}^{\prime }$\textbf{\ and }$\mathbf{1s2lnl}%
^{\prime }$\textbf{\ of highly charged ions with }$\mathbf{Z=6-36}$\textbf{. 
}\\
\textbf{I. Transitions from the states with }$\mathbf{n=2,3}$.\\
\\
}

\begin{abstract}
The wavelengths and radiative transition probabilities for transitions $%
2lnl^{\prime }-1snl^{\prime \prime },2lnl^{\prime }-1s2l^{\prime \prime
},1s2lnl^{\prime }-1s^{2}nl^{\prime \prime },1s2lnl^{\prime
}-1s^{2}2l^{\prime \prime },$ and the autoionization decay probabilities for
doubly excited states $2lnl^{\prime },1s2lnl^{\prime }$ were calculated in
ions with atomic numbers $Z=6-36$ for $n=2-10,$ $l^{\prime }=0-3.$ The
calculations were carried out by means of the MZ code based on the
Z-expansion method. Relativistic corrections were taken into account within
the framework of the Breit operator. The main difference with previous
calculations by MZ code consists in accounting for the first order
corrections in powers of 1/Z, corresponding to the screening effects, in
calculations of autoionization rates. New data for comparatively large rates
are about 20-50\% less as compared to previous ones and are in a agreement
within 10\% with the results of calculations made by the methods based on
multi-configuration wave functions with non-relativistic and relativistic
orbitals. Some refinements and corrections concerning the energies and
radiative transition probabilities were also introduced in the MZ code. In
this paper the main formulas used in a modified MZ-code and the data needed
for description of dielectronic satellites with $n=2,3$ are given; the data
for higher $n$ will be\ presented in the following publications.
\end{abstract}

\section{Introduction}

Dielectronic satellite (DS) spectra in the vicinity of the resonance lines
of [H] and [He] ions provide the important information on electron
temperature, density, ionization state and other characteristics of hot
plasmas and are widely used in spectroscopic diagnostics. For unambiguous
analysis and modeling of DS spectra one usually needs to have a large amount
of atomic data: wavelengths, radiative transition probabilities and
autoionization rates, calculated with high accuracy. The approach of
expansion over 1/Z (Z is the nuclear charge), called also "Z-expansion"
method, is an effective one for determination of atomic characteristics for
highly charged ions. MZ code based on this approach was developed for the
calculations of wavelengths and transition probabilities. The input
information (coefficients of the Z-expansion) can be used for an entire
isoelectronic sequence and total calculation time on the modern PC is less
than 1 min. However the coefficients of the Z-expansion must be calculated
separately.

The outline of Z-expansion method and the calculations of corresponding data
for DS emitted by [He] and [Li] ions are given in the \cite{PREP6, PREP146,
PREP188, VS78, PREP212} and by [Be] through [Ne] ions in the papers \cite%
{PREP58, JPhysB}. The description of the MZ code and corresponding formulas
are given in the monograph \cite{BOOK}. The systematic calculations of
atomic data for doubly excited states $2lnl^{\prime }$ and $1s2lnl^{\prime }$
of ions with $Z=6-33$ by this code have been performed and published in the
late seventies for $n=2$ \cite{VS78} and $n=3$ \cite{VS80}.

The comparison of the wavelengths calculated by Z-expansion method with the
experimental data obtained from the tokamak and EBIT K-spectra of [Li] argon
ions \cite{BITTER, MARCH} has shown that this approach provides better fit
to the measured values than the calculations, employing the MCHF
(multi-configuration Hartree-Fock) \cite{SUPST} or MCDF (Dirac-Fock) method 
\cite{CHEN, NILS}. The radiative probabilities obtained up to the second
term over 1/Z have been found in agreement with the calculations of \cite%
{SUPST, CHEN, NILS} within 10\%, while the autoionization rates, obtained in
the first nonvanishing order, differed up to a factor\ of 2.

The comparison of the satellite lines of [H] and [He] spectra calculated by
the MZ and AUTOLSJ methods for Fe, Ca and S ions is given in \cite{KATO}.

In this work the calculations have been carried out by means of a modified
version of MZ code, completed with some corrections and refinements. In
particular, the second order corrections to the autoionization radial
integrals, corresponding to the screening effects, as well as the
contribution of cascade transitions (3-2) to the intensity of DS have been
included. The wavelengths and radiative transition probabilities for
transitions $2lnl^{\prime }-1snl^{\prime \prime },$ $2lnl^{\prime
}-1s2l^{\prime \prime },$\ $1s2lnl^{\prime }-1s^{2}nl^{\prime \prime },$\ $%
1s2lnl^{\prime }-1s^{2}2l^{\prime \prime }$ and the autoionization decay
probabilities for doubly excited states $2lnl^{\prime },$ $1s2lnl^{\prime }$
have been calculated in ions with atomic numbers $Z=6-36$ for $n=2-10,$ $%
l^{\prime }=0-3.$ The present paper contains the main formulas used in a
modified MZ-code and the data for dielectronic satellites with $n=2,3$; the
data for higher $n$ will be\ given in the following publications.

Hereafter the atomic units and the following notations are used 
\begin{equation}
\begin{array}{c}
\left[ j_{1}j_{2}\;...\right] =\left( 2j_{1}+1\right) ^{1/2}\left(
2j_{2}+1\right) ^{1/2}...\smallskip \  \\ 
\delta \left( j_{1}j_{2}\;...,j_{1}^{\prime }j_{2}^{\prime }\;...\right)
=\delta \left( j_{1},j_{1}^{\prime }\right) \delta \left(
j_{2},j_{2}^{\prime }\right) \;...\smallskip \ 
\end{array}
\label{1.1}
\end{equation}%
$\alpha =e^{2}/\hbar c=1/137.0362,\;\;Ry=e^{2}/2a_{0}=109737\;cm^{-1}.$

An isoelectronic sequence of an atom A is designated as [A]; the number of
electrons and the nuclear charge are denoted through $N$ and $Z$,
respectively.

The atomic system is divided into two groups: the group of the outer
electrons $\gamma _{n}$ ($\nu $ electrons with the principal quantum number $%
n$) and the core $\gamma _{c}$ with $N_{c}=N-\nu $ electrons. The ionic
states are described by the set of atomic numbers 
\begin{equation}
aJ=\gamma _{c}(S_{c}L_{c})\gamma _{n}(S_{n}L_{n})SLJ  \label{1.2}
\end{equation}

Below in formulas for angular factors we use the $\mathcal{M}$ factors
defined as:%
\begin{equation}
\mathcal{M}_{r_{3}r_{2}r_{1}}(j_{3}j_{2}j_{1})=\mathcal{M}%
_{r_{3}r_{2}r_{1}}(j_{3}j_{2}j_{1},j_{3}^{\prime }j_{2}^{\prime
}j_{1}^{\prime })=\left[ j_{1}j_{1}^{\prime }r_{1}\right] \left\{ 
\begin{array}{ccc}
j_{3} & j_{3}^{\prime } & r_{3} \\ 
j_{2} & j_{2}^{\prime } & r_{2} \\ 
j_{1} & j_{1}^{\prime } & r_{1}%
\end{array}%
\right\} .  \label{1.3}
\end{equation}

\section{Method of\textbf{\ calculation}}

\subsection{Energies}

Z-expansion method is based on the perturbation theory expansion over 1/Z (Z
is the nuclear charge), in other words, on the hydrogen-like basis. The
relativistic corrections are included in the framework of the Breit
Hamiltonian.

The energy matrix $E$ for a given Layzer complex (states with given $J$,
parity and set of principal quantum numbers of electrons) can be presented
in the form

\begin{equation}
E(aJ,a^{\prime }J)=[E^{N}+E^{R}]\delta (SL,S^{\prime }L^{\prime
})+E^{S}+\Delta E^{x}\delta (a,a^{\prime })  \label{2.1}
\end{equation}%
Here the matrices $E^{N},E^{R}$ and $E^{S}$ correspond to the
nonrelativistic energy, spin independent part and spin dependent part of the
Breit operator, respectively. The term $\Delta E^{x}$ is introduced for
additional semi-empirical correction. The matrix $E^{N}$ includes the terms
up to the second order and the matrices $E^{R}$ and $E^{S}$ contain the zero
and first orders (over electrostatic interaction of electrons): 
\begin{equation}
\begin{array}{c}
E^{N}=Z^{2}E_{0}\delta \left( a,a^{\prime }\right) +ZE_{1}+E_{2}\smallskip \ 
\\ 
E^{R}=\frac{1}{4}\alpha ^{2}Z^{3}\left[ Z\rho _{0}\delta \left( a,a^{\prime
}\right) +\rho _{1}+\rho ^{\prime }\right] \smallskip \  \\ 
E^{S}=\frac{1}{4}\alpha ^{2}Z^{3}\left[ \left( Z\epsilon _{0}+\epsilon
_{1}+\epsilon ^{\prime }\right) Q_{1}(J)+\epsilon ^{\prime \prime }Q_{2}(J)%
\right]%
\end{array}
\label{2.2}
\end{equation}

The terms containing $\rho _{0},\epsilon _{0}$ and $\rho _{1},\epsilon _{1}$
give the contributions of the zero and first orders from one-particle terms
of the Breit operator. The terms including $\rho ^{\prime },\epsilon
^{\prime }$ and $\epsilon ^{\prime \prime }$ correspond to the zero order
contributions from two-particle terms. The coefficients $Q_{r}$ reflect the
tensor character of the rank $r$ of the interactions in $E^{S}$: 
\begin{equation}
Q_{r}(J)\equiv Q_{r}(SLJ)=(-1)^{r}\frac{1}{[J]}\mathcal{M}_{rr0}\left(
SLJ\right) =(-1)^{S+L^{\prime }+J}\left\{ 
\begin{array}{ccc}
S & L & J \\ 
L^{\prime } & S^{\prime } & r%
\end{array}%
\right\}  \label{2.3}
\end{equation}

We found that it is useful to rewrite the formulas for the diagonal matrix
elements in the "screening" representation:

\begin{eqnarray}
E^{N} &=&(Z-\sigma )^{2}E_{0}+\widetilde{E}_{2}\smallskip ,\ \ \sigma =-%
\dfrac{E_{1}}{2E_{0}},\ \ \widetilde{E}_{2}\smallskip =E_{2}-\sigma ^{2}E_{0}
\label{2.5} \\
E^{R} &=&\frac{1}{4}\alpha ^{2}\left( Z-\sigma ^{R}\right) ^{3}\left( Z\rho
_{0}+\rho ^{\prime }\right) ,\ \ \sigma ^{R}=-\dfrac{\rho _{1}}{3\rho _{0}}%
\smallskip \;\ \ \ \ \ \ \text{for }\;a=a^{\prime }
\end{eqnarray}%
\begin{equation}
\text{ \ \ \ \ \ }E^{N}=ZE_{1}+E_{2},\;\;E^{R}=\frac{1}{4}\alpha
^{2}Z^{3}(\rho _{1}+\rho ^{\prime })\ \;\ \ \ \ \ \ \ \ \ \ \ \ \ \text{for\
\ }a\neq a^{\prime }  \label{2.6}
\end{equation}%
For $E^{N}$\ Eq.(\ref{2.5}) gives the same result as Eq.(\ref{2.2}), however
in most cases $\left\vert \widetilde{E}_{2}\smallskip \right\vert \ll
\left\vert E_{2}\right\vert $, meaning that the term $\sigma ^{2}E_{0}$
describes a significant part of the second order term $E_{2}$.\ For $E^{R}$\
the Eqs.(\ref{2.5}) and (\ref{2.2}) provide different results since the 2'd
order initially was not included in $E^{R}$. Numerical calculations with a
screening representation (\ref{2.5}) are in a better agreement with the
experimental data.

For $E^{S}$, due to its tensor nature, nondiagonal zero order matrix element 
$\epsilon _{0}$ does not vanish\ and the screening representation

\begin{equation}
E^{S}=\frac{1}{4}\alpha ^{2}\left[ \left( Z-\sigma ^{S}\right)
^{3}(Z\epsilon _{0}q_{1}+\varepsilon ^{\prime })Q_{1}(J)+\varepsilon
^{\prime \prime }Q_{2}(J)\right]  \label{2.7}
\end{equation}%
takes place for $a=a^{\prime }$ and $a\neq a^{\prime }$. The additional
angular factors $q$ are independent on $J$ (for details see \cite{BOOK}).

Note that in the MZ code while employing the screening representation the
formulas similar to (\ref{2.5}-\ref{2.7}) have been used but written
separately \ for the outer electrons $\gamma _{n}$ and the core $c$ with 
\begin{equation}
E_{0}(\gamma _{n})=-\nu /n^{2},\ \ E_{0}(c)=-\sum_{i}\nu _{i}/n_{i}^{2},\ \
\sum_{i}\nu _{i}=N_{c}  \label{2.6a}
\end{equation}

Coefficients $E$, $\rho $ and $\varepsilon $ were calculated by the
perturbation theory method with the hydrogen-like basis. Eigenvalues $%
\widetilde{E}_{J}$ and eigenvectors $C_{J}$ were determined by
diagonalization of the energy matrix (\ref{2.1}). The corrections $E^{L}$=$%
O(\alpha ^{3})$ as well as due to nuclear mass were then introduced to the
eigenvalue of energy:

\begin{equation}
E_{J}=[\widetilde{E}_{J}+\Delta E^{L}]\cdot (1+\frac{1}{1836M}),\ \ M\approx
2Z.  \label{3.1}
\end{equation}

The term $\Delta E^{L}$ can be written as

\begin{equation}
\Delta E^{L}=\frac{4}{3\pi }\alpha ^{3}Z_{1}^{4}\Lambda +\alpha
^{4}Z_{1}^{6}D,\qquad Z_{1}=Z-\sigma ^{L}  \label{3.2}
\end{equation}%
The second term gives the correction due to the difference of the Dirac and
Pauli formulas and are expressed through the one-electron contributions
equal to:%
\begin{equation}
\begin{array}{c}
x^{4}D(nlj)=-\frac{1}{2}\left\{ \frac{1}{NN^{\prime }}-\frac{1}{n^{2}}+\frac{%
x^{2}}{n^{3}}\left( \frac{2}{2j+1}-\frac{3}{4n}\right) \right\} \smallskip
\\ 
N=\left[ n^{2}-2(n-k)\delta \right] ^{1/2},\ \ N^{\prime }=\frac{1}{2}\left(
N+n-\delta \right) \\ 
\delta =k-\left( k^{2}-x^{2}\right) ^{1/2},\ \ x=\alpha Z_{1},\ \ k=j+\frac{1%
}{2}%
\end{array}
\label{3.3}
\end{equation}%
The first term in (\ref{3.2}) gives the correction for the Lamb-shift
including vacuum polarization. It is described by the rather complicated
semi-empirical formula (see \cite{BOOK}).

\subsection{Transition probabilities}

\textbf{The radiative transition probability} $A(aJ,a^{\prime }J^{\prime })$
is given by%
\begin{equation}
A(aJ,a^{\prime }J^{\prime })=\dfrac{A_{\kappa }}{2J+1}[E_{J}(a)-E_{J^{\prime
}}(a^{\prime })]^{3}\medskip \left\vert
\sum\limits_{a_{1}a_{2}}C_{J}(a,a_{1})T_{12}C_{J^{\prime }}(a_{2},a^{\prime
})\right\vert ^{2}  \label{4}
\end{equation}%
where $C_{J}$ are the eigenvectors obtained by the diagonalization of the
energy matrix. For the dipole transition $\kappa =1$ and $A_{1}=4c\alpha
^{4}/3a_{0}$. The transition matrix $T_{12}$ is calculated up to terms $1/Z$:%
\begin{equation}
\begin{array}{c}
T_{12}=T_{12}(a_{1}J_{1},a_{2}J_{2})=T^{(0)}+{\frac{1}{Z}}T^{(1)}\smallskip
\\ 
T^{(0)}=B_{\kappa }\left( a_{1}J_{1},a_{2}J_{2}\right) R_{\kappa }\left(
n_{1}l_{1},n_{2}l_{2}\right) \smallskip \\ 
R_{\kappa }\left( n_{1}l_{1},n_{2}l_{2}\right) =[l_{1}l_{2}]\left( 
\begin{array}{ccc}
l_{1} & l_{2} & \kappa \\ 
0 & 0 & 0%
\end{array}%
\right) \int\limits_{0}^{\infty }P_{1}(r)P_{2}(r)r^{\kappa }dr%
\end{array}
\label{4.1}
\end{equation}%
where $B_{\kappa }$ and $R_{\kappa }$\ are the angular and radial parts of
the $\kappa $-pole matrix element, respectively (see \cite{BOOK}); $\
P_{i}(r)$ are the hydrogen-like (Coulomb) radial functions.\medskip

\textbf{The probability of autoionization }decay of the ion $X_{N}(aJ)$ to $%
X_{N-1}(\alpha _{0}J_{0})$ with emission of the electron $kl$ is described
by a more complicated expression. This is due to the fact that the initial
and final N-electron states are described in different coupling schemes: $%
a=\alpha _{p}nlSL,\ \ a_{0}=\alpha _{0}J_{0}.kslj$. However, assuming the
absence of a configuration mixing for $X_{N-1}(\alpha _{0}J_{0})$, one can
sum up the probability over $J_{0}lj$ to obtain a simpler formula sufficient
in most practical cases: 
\begin{equation}
\begin{array}{c}
\Gamma (aJ,\alpha _{0})=\Gamma _{0}\sum_{lS_{1}L_{1}}\left\vert
\sum\limits_{a_{1}}C_{J}\left( a,a_{1}\right) \gamma \left(
a_{1},a_{2}\right) \right\vert ^{2} \\ 
\Gamma _{0}=2Ry/\hbar =4.1347\cdot 10^{16}c^{-1},\ \ a_{0}=\alpha
_{0}k_{0}l_{0},\ \ k_{0}^{2}/2=E_{J}(a)-E(\alpha _{0}).%
\end{array}
\label{6}
\end{equation}%
Here $a_{2}=\alpha _{0}klS_{2}L_{2}$ is a final N-electron state in the
"usual coupling scheme", $k_{0}^{2}/2$ and $l$ are the energy and the
orbital momentum of the ejected electron, respectively, and $\gamma
(a_{1},a_{0})$ is the decay amplitude. In the zero-order the decay
amplitudes are equal: 
\begin{equation}
\begin{array}{c}
\gamma (a_{1},a_{2})=\sum_{\kappa ^{\prime }}b_{\kappa ^{\prime }}R_{\kappa
^{\prime }}(l_{1}l_{2},l^{\prime }l_{0})+\Sigma _{\kappa ^{\prime \prime
}}g_{\kappa ^{\prime \prime }}R_{\kappa ^{\prime \prime
}}(l_{1}l_{2},l_{0}l^{\prime })\smallskip \  \\ 
R_{\kappa }(l_{1}l_{2},l_{4}l_{3})=[l_{1}l_{2}l_{3}l_{4}]\left( 
\begin{array}{ccc}
l_{1} & l_{3} & \kappa \\ 
0 & 0 & 0%
\end{array}%
\right) \left( 
\begin{array}{ccc}
l_{2} & l_{4} & \kappa \\ 
0 & 0 & 0%
\end{array}%
\right) \smallskip \  \\ 
\times \int\limits_{0}^{\infty }P_{1}(r)P_{2}(r^{\prime })\frac{%
r_{<}^{\kappa }}{r_{>}^{\kappa +1}}P_{4}(r^{\prime })P_{3}(r)dr^{\prime }dr%
\end{array}
\label{7}
\end{equation}%
where the hydrogen-like radial functions are used for $P_{i}(r).$ The
continuum functions are normalized as

\begin{equation}
P_{kl}\simeq k^{-1/2}\sin (kr-\frac{l\pi }{2}+\frac{1}{k}\ln (2kr)+\delta
_{l})  \label{4.13}
\end{equation}%
Some formulas for the angular parts from Eqs.(\ref{4.1}, \ref{7}) are given
in the Appendix.

In the previous papers \cite{VS78}, \cite{VS80} the zero-order approximation
was used for the decay amplitudes. In the present work we added the
corrections of the first order for the radial integrals in the decay
amplitude:%
\begin{equation}
R_{\kappa }=R_{\kappa }^{(0)}+\frac{Z_{0}}{Z}\Delta R_{\kappa },\ \ \ \Delta
R_{\kappa }=R_{\kappa }^{(a)}(Z_{0})-R_{\kappa }^{(0)}  \label{9}
\end{equation}%
where $R_{\kappa }^{(0)}$ are\ the radial integrals defined in (\ref{7}) and 
$R_{\kappa }^{(a)}\ $are the similar integrals but calculated with the
radial functions obtained in the effective central field for the ion with $%
Z=Z_{0}$ by the code ATOM \cite{BOOK} ($Z_{0}=18$ was used in the present
work). This procedure evidently takes into account only the screening
effects in the 1'st order corrections to $\gamma (a_{1},a_{0})$.

\subsection{ Intensity of dielectronic satellites}

The intensity of the dielectronic satellite line $aJ-a^{\prime }J^{\prime }$
is proportional to the factor

\begin{equation}
Q_{d}(aJ-a^{\prime }J^{\prime })=\dfrac{g_{aJ}A(aJ-a^{\prime }J^{\prime
})\Gamma (aJ-\alpha _{0}J_{0})}{A(aJ)+\Gamma (aJ)}=K(aJ-a^{\prime }J^{\prime
})g_{aJ}\Gamma (aJ-\alpha _{0}J_{0})  \label{a61}
\end{equation}%
where the branching ratio $K(aJ-a^{\prime }J^{\prime })$ is given by%
\begin{equation}
K(aJ-a^{\prime }J^{\prime })=\dfrac{A(aJ-a^{\prime }J^{\prime })}{%
A(aJ)+\Gamma (aJ)}  \label{a62}
\end{equation}%
and 
\begin{equation}
A(aJ)=\sum_{a"J"}A(aJ-a"J"),\ \ \ \Gamma (aJ)=\sum_{a"J"}\Gamma (aJ-\alpha
_{0}"J_{0}")  \label{a63}
\end{equation}%
are total probabilities of the radiative and autoionization decays, $%
g_{aJ}=2J+1$ is the statistical weight, $\alpha _{0}J_{0}$ is the state of
the initial ion $X_{Z}$ for the dielectronic recombination ($1s^{2}$ for
[Li] satellites and $1s$ for [He] satellites). The intensity depends also on
electron temperature $T$ through the factor $exp[-(\Delta E-\delta E)/T]$,
where $\Delta E$ is the excitation energy of the main transition (resonance
or higher member of a spectral sequence) in the ion $X_{Z}$, $\delta E$ is
the bound energy of the captured electron. In most cases $\delta E\ll \Delta 
\dot{E}$.

Cascade transitions ($n=3\rightarrow n=2)$ may increase the total intensity
of the $2-1$ satellites and decrease the intensity of the $3-1$ satellite.
The latter effect is accounted for by including the states $n"=2$ in the sum
(\ref{a63}). The relative contribution of cascades to the intensity of the $%
2-1$ satellites generally depends on $T$ as $exp[-\delta E_{32}/T]$. In the
limit $T\gg \delta E_{32}$ the total intensity is proportional to the factor 
$Q_{d}^{t}:$%
\begin{equation}
\begin{array}{c}
Q_{d}^{t}(aJ-a^{\prime }J^{\prime })=Q_{d}(aJ-a^{\prime }J^{\prime
})+K(aJ-a^{\prime }J^{\prime })\cdot \sum_{a"J"}Q_{d}(a"J"-aJ)\smallskip \\ 
=Q_{d}(aJ-a^{\prime }J^{\prime })\cdot \left( 1+\Delta _{c}\right) ,\ \ \
\Delta _{c}=\sum\limits_{a"J"}\dfrac{g_{a"J"}\Gamma (a"J"-\alpha _{0}J_{0})}{%
g_{aJ}\Gamma (aJ-\alpha _{0}J_{0})}K(a"J"-aJ)%
\end{array}
\label{a64}
\end{equation}%
where $a"$ are the levels with $n=3$.

\section{\textbf{EXPLANATION OF TABLES}}

In the Tables Ia, IIa, Ib, IIb the results of calculations are given for
atomic data connected with the transitions from doubly excited states of
He-like (I,IIa) and Li-like (I,IIb) ions with Z=6-36. Data for resonance
lines of [H] and [He] ions as well as for intercombination,
magneto-quadrupole and forbidden lines of [He] ion are also included. The
notations as well as the format of these data are similar to those used in
the papers \cite{VS78}, \cite{VS80}. The set of quantum numbers describing
the initial state is denoted through $\gamma $=($aSLJ$).

The Tables Ia, IIa, Ib, IIb contain for each transition $\gamma \rightarrow
\gamma ^{\prime }$ the following data:

$\Delta E$ is transition energy in keV,

$\lambda $ - wavelength in Angstroms,

$A=A(\gamma ,\gamma ^{\prime })$ - probability of radiative decay, Eq.(\ref%
{4}), in units 10$^{13}$ $s^{-1}$,

$\Gamma (\gamma )$ - probability of autoionization, Eq.(\ref{6}), in units 10%
$^{13}$ $s^{-1}$,

$K$ - branching ratio, Eq.(\ref{a62}),

$Q_{d}$ - satellite intensity factor, Eq.(\ref{a61}-\ref{a63}), in units 10$%
^{13}$ $s^{-1}$.

To avoid a confusion, it is worth to give here the correspondence with the
notations used in some papers quoted above: $A=A^{r}$, $\Gamma =A^{a},$ $%
Q_{d}=g_{0}F_{2}.$ Here $g_{0}$ is the statistical weight of the initial
state from which DS is excited ($1s$ or $1s^{2}$).

In the Tables these data are denoted as follows.

$\Delta E$ = E , $\lambda $ = WL, $\Gamma $ = G.

The set of quantum numbers $a$ (electronic configuration and intermediate
quantum numbers specifying the coupling scheme for angular momenta) is
denoted through the capital letters:

\bigskip

I. He-like ions

\bigskip $%
\begin{array}{ccc}
\text{Y=1s} & \text{C=2s2p} & \text{A'=2p3d} \\ 
\text{R=2p} & \text{E=2s2s} & \text{B'=2p3s} \\ 
\text{R'=3p} & \text{F=2p2p} & \text{C'=2s3p} \\ 
& \text{P=1s2p} & \text{F'=2p3p} \\ 
& \text{S=1s2s} & \text{G'=2s3d} \\ 
& \text{S'=1s3s} & \text{E'=2s3s} \\ 
& \text{P'=1s3p} & \text{D'=1s3d}%
\end{array}%
$

\ \ \ \ \ \ \ \ \ \ \ \ \ \ \ \ \ \ \ \ \ \ \ \ \ 

II. Li-like ions

\bigskip \bigskip 
\begin{tabular}{ccc}
Y=1s$^{2}$ & C=2s2p($^{1}$P)1s & A'=1s2p($^{1}$P)3d \\ 
R=1s2p & E=2s$^{2}$($^{1}$S)1s & B'=1s2p($^{1}$P)3s \\ 
R'=1s3p & F=2p$^{2}$($^{1}$S)1s & C'=1s2s($^{1}$S)3p \\ 
& F=2p$^{2}$($^{1}$D)1s & F'=1s2p($^{1}$P)3p \\ 
& M=2p$^{2}$($^{3}$P)1s & G'=1s2s($^{1}$S)3d \\ 
& K=2s2p($^{3}$P)1s & E'=1s2s($^{1}$S)3s \\ 
& S=1s$^{2}$2s & I'=1s2p($^{3}$P)3d \\ 
& P=1s$^{2}$2p & J'=1s2p($^{3}$P)3s \\ 
& S'=1s$^{2}$3s & K'=1s2s($^{3}$S)3p \\ 
& P'=1s$^{2}$3p & L'=1s2s($^{3}$S)3s \\ 
& D'=1s$^{2}$3d & M'=1s2p($^{3}$P)3p \\ 
&  & N'=1s2s($^{3}$S)3d%
\end{tabular}

In the Tables below the numbers following the letters mean the statistical
weights (2S+1) (2L+1) (2J+1). For example: R234-Y212 stands for 2p$^{2}$P$%
_{3/2}$-1s$^{2}$S$_{1/2}$ and A'353-D'155 means 2p3d$^{3}$D$_{1}$-1s3d$^{1}$D%
$_{2}$.

In conclusion we would like to note that the atomic data presented in Tables
Ia, IIa and Ib, IIb could be provided in another formats (e.g. convinient
for Matlab package) due to request by the adress \textbf{urnov@sci.lebedev.ru%
} or \textbf{vainsh@sci.lebedev.ru}.

\bigskip\ 

\section{Acknowlegments}

The work was supported by the Grants of Russian Foundation for Baisic
Research 03-02-16053 and 06-02-16298-a.

\bigskip

\appendix\pagebreak

\section{\textbf{Some additional formulas}}

\subsection{Zero order energies}

The one electron zero order coefficients in Eqs.(\ref{2.5},\ref{2.7}) are
equal to: 
\begin{equation}
\begin{array}{c}
E_{0}\left( nl\right) =-\dfrac{1}{2n^{2}},\ \ \smallskip \ \rho _{0}\left(
nl\right) =-\dfrac{2}{n^{3}}\left[ 2/\left( 2l+1\right) -3/4n-\delta \left(
l,0\right) \right] \\ 
\epsilon _{0}(nl)=\dfrac{2}{n^{3}}\left[ \dfrac{6}{l\left( l+1\right) \left(
2l+1\right) }\right] ^{1/2}\;\left( l\neq 0\right) ,\;\;\;\epsilon
_{0}\left( nl\right) =0\;\;\left( l=0\right)%
\end{array}
\label{a1}
\end{equation}%
The Pauly one electron energy is:%
\begin{equation}
\varepsilon ^{P}(nlj)=E_{0}+Z^{2}\dfrac{\alpha ^{2}}{4}\left( \rho
_{0}+\epsilon _{0}Q_{1}\right) =-\dfrac{1}{2n^{2}}-Z^{2}\dfrac{\alpha ^{2}}{%
2n^{3}}\left( \frac{2}{2j+1}-\frac{3}{4n}\right)  \label{a2}
\end{equation}%
The corresponding Dirac energy can be written as:%
\begin{equation}
\begin{array}{c}
\varepsilon ^{D}=\varepsilon ^{P}+x^{4}D=-\dfrac{1}{2NN^{\prime }},\ \ N=%
\left[ n^{2}-2(n-k)\delta \right] ^{1/2},\ \ N^{\prime }=\left( N+n-\delta
\right) /2 \\ 
\delta =k-(k^{2}-x^{2})^{1/2},\ \ k=j+\frac{1}{2},\ \ x=\alpha Z%
\end{array}
\label{a3}
\end{equation}%
In the screening representation $Z$ should be replaced by $Z_{1}=Z-\sigma
^{L}\approx Z-\sigma ^{S}$.

\subsection{Angular parts of transition matrix elements}

\paragraph{Radiative transitions}

Below we use the symmetrical over spin and orbital variables designations.
Greek letters are used for pairs of orbital and spin momenta or multipoles:

\begin{equation}
\tau =SL,\;\lambda =sl,\;\kappa =qk,\;s=\frac{1}{2}  \label{a41}
\end{equation}%
Functions with Greek arguments in angular parts are in fact the products of
the similar orbital and spin factors:

\begin{equation}
F(\tau \lambda \kappa )=F(Ssq)\cdot F(Llk)  \label{a42}
\end{equation}

The angular part of the matrix $T^{0}=B_{\kappa }\left(
a_{1}J_{1},a_{2}J_{2}\right) R_{\kappa }\left( n_{1}l_{1},n_{2}l_{2}\right) $
for the simplest case $a=S_{c}L_{c}nlSL$ is equal to:%
\begin{equation}
\begin{array}{c}
S_{c}L_{c}n_{1}l_{1}S_{1}L_{1}J_{1}-S_{c}L_{c}n_{2}l_{2}S_{2}L_{2}J_{2}%
\smallskip \\ 
B_{\kappa }\left( a_{1}J_{1},a_{2}J_{2}\right) =B_{\kappa }(\tau )\mathcal{M}%
_{0kk}(SLJ)\smallskip \\ 
B_{\kappa }(\tau )=B_{\kappa }^{0}(\tau )\equiv \lbrack \tau _{c}]\mathcal{M}%
_{0\kappa \kappa }(\tau _{c}\lambda \tau )=\frac{[S]}{2}\delta (S,S^{\prime
})[L_{c}]\mathcal{M}_{0kk}(L_{c}lL)\smallskip%
\end{array}
\label{a4}
\end{equation}%
for transitions from the shell of equivalent electrons $\lambda ^{m}$: 
\begin{equation}
\lambda ^{m}\tau -\lambda ^{m-1}(\tau _{p})\lambda ^{\prime }\tau ^{\prime
},\;\;\;B_{\kappa }\left( \tau \right) =\sqrt{m}G_{\tau _{p}}^{\tau
}B_{\kappa }^{0}\left( \tau _{p}\lambda \tau \right)  \label{2.113}
\end{equation}%
and for transitions from the inner shell $\lambda ^{m}$ without or with the
change of the coupling: 
\begin{equation}
\begin{array}{c}
\lambda ^{m}(\tau _{2})\tau _{1}\tau -\lambda ^{m-1}(\tau _{c})\tau
_{1}(\tau _{p})\lambda ^{\prime }\tau ^{\prime },\ \ B_{\kappa }\left( \tau
\right) =\sqrt{m}G_{\tau _{c}}^{\tau _{2}}B_{\kappa }^{0}\left( \tau
_{p}\lambda \tau \right) C_{1}\smallskip \\ 
\lambda ^{m}(\tau _{2})\tau _{1}\tau -\lambda ^{m-1}(\tau _{c}),\tau
_{1}\lambda ^{\prime }(\tau _{p})\tau ^{\prime },\ \ B_{\kappa }\left( \tau
\right) =\sqrt{m}G_{\tau _{c}}^{\tau _{2}}B_{\kappa }^{0}\left( \tau
_{p}\lambda \tau \right) C_{2}\smallskip \\ 
C_{1}=(\tau _{c}\lambda (\tau _{2})\tau _{1}\tau \ |\ \tau _{c}\tau
_{1}(\tau _{p})\lambda \tau ),\ \ C_{2}=(\tau _{c}\lambda (\tau _{2})\tau
_{1}\tau \ |\ \tau _{c},\tau _{1}\lambda (\tau _{p})\tau )\smallskip \\ 
m=N=[\lambda ]^{2}:\ \tau _{2}=0,\ C_{1}=C_{2}=\frac{[\tau _{p}]}{[\tau
\lambda ]}\smallskip \\ 
m=1:\ \tau _{c}=0,\;\tau _{2}=\lambda ,\ C_{1}=C_{2}=1%
\end{array}
\label{2.113a}
\end{equation}

\paragraph{Autoionization transitions}

The angular factors $b_{\kappa }$ and $g_{\kappa }$ depend on transition $%
a_{1}-a_{2}$. For autoionizing states $l_{1}l_{2}$ and $1sl_{1}l_{2}$ one
has: 
\begin{equation}
\begin{array}{c}
b_{\kappa ^{\prime }}=C^{\prime }(-1)^{l_{2}}[l^{\prime }l_{1}]^{-1}\delta
(\kappa ,l_{1}),\;\;g_{\kappa ^{\prime \prime }}=C^{\prime \prime
}(-1)^{l_{1}}[l^{\prime }l_{2}]^{-1}\delta (\kappa "l_{2})_{:}\smallskip \\ 
l_{1}l_{2}-1skl^{\prime }:\;\;C^{\prime }=1,\;C^{\prime \prime
}=(-1)^{s}\cdot 2\smallskip \\ 
1sl_{1}(l_{1}S_{p})l_{2}-1s^{2}kl^{\prime }:\;C^{\prime }=\delta
(S_{p},0),\;C^{\prime \prime }=-1\smallskip \\ 
l_{1}l_{2}(L_{1}S_{p})1s-kl^{\prime }:\;\;C^{\prime }=(-1)^{S_{p}-1}\frac{%
\left[ S_{p}\right] }{2},\;C^{\prime \prime }=-\frac{\left[ S_{p}\right] }{2}%
\end{array}
\label{4.14}
\end{equation}

\newpage
\normalsize
Table Ia. Transitions 2l'2l"-1s2l in [He] and 2p-1s in [H] ions
\scriptsize
%TCIDATA{Version=5.00.0.2606}
%TCIDATA{LaTeXparent=0,0,arxiv.tex}

%%%\scriptsize

\begin{center}

\begin{longtable}{cccccccr}
%%% \caption{Table caption} \\

\hline
\hline
  WL, \AA & E, keV & A & G & K & QD & Transition & No \\
\hline
\endfirsthead

\hline
  WL, \AA & E, keV & A & G & K & QD & Transition & No \\
\hline
\endhead

\hline
\multicolumn{8}{r}{Continued on next page} \\
\hline
\endfoot

\hline \hline
\endlastfoot

\multicolumn{8}{l}{$  n= 2;  Z= 6;$} \\
$33.569$ & $.36934$ & $9.44(-08)$ & $5.10(-01)$ & $1.53(-07)$ & $7.7859(-08)$ & {\tt F\ 111-P\ 333} & $1$ \\
$33.569$ & $.36934$ & $0.00(+00)$ & $5.10(-01)$ & $0.00(+00)$ & $0.0000(+00)$ & {\tt F\ 111-P\ 331} & $2$ \\
$33.734$ & $.36753$ & $8.14(-02)$ & $0.00(+00)$ & $1.00(+00)$ & $0.0000(+00)$ & {\tt R\ 234-Y\ 212} & $3$ \\
$33.740$ & $.36748$ & $8.13(-02)$ & $0.00(+00)$ & $1.00(+00)$ & $0.0000(+00)$ & {\tt R\ 232-Y\ 212} & $4$ \\
$33.805$ & $.36677$ & $8.17(-07)$ & $1.44(+01)$ & $5.66(-08)$ & $2.4394(-06)$ & {\tt C\ 133-S\ 313} & $5$ \\
 \\
$33.890$ & $.36585$ & $1.08(-01)$ & $5.10(-01)$ & $1.75(-01)$ & $8.9330(-02)$ & {\tt F\ 111-P\ 133} & $6$ \\
\end{longtable}

\end{center}

%%%\end{document}

\newpage
\normalsize
Table Ib. Transitions 1s2l'2l"-1s2.2l in [Li] and 1s2p-1s2 in [He] ions 
\scriptsize
%TCIDATA{Version=5.00.0.2606}
%TCIDATA{LaTeXparent=0,0,arxiv.tex}

%%%\scriptsize

\begin{center}

\begin{longtable}{cccccccr}
%%% \caption{Table caption} \\

\hline
\hline
  WL, \AA & E, keV & A & G & K & QD & Transition & No \\
\hline
\endfirsthead

\hline
  WL, \AA & E, keV & A & G & K & QD & Transition & No \\
\hline
\endhead

\hline
\multicolumn{8}{r}{Continued on next page} \\
\hline
\endfoot

\hline \hline
\endlastfoot

\multicolumn{8}{l}{$  n= 2;  Z= 6;$} \\
$40.281$ & $.30780$ & $9.07(-02)$ & $0.00(+00)$ & $1.00(+00)$ & $0.0000(+00)$ & {\tt R\ 133-Y\ 111} & $1$ \\
$40.710$ & $.30456$ & $1.13(-02)$ & $4.48(-01)$ & $2.35(-02)$ & $2.1016(-02)$ & {\tt F\ 212-P\ 232} & $2$ \\
$40.712$ & $.30454$ & $2.31(-02)$ & $4.48(-01)$ & $4.79(-02)$ & $4.2908(-02)$ & {\tt F\ 212-P\ 234} & $3$ \\
$40.716$ & $.30451$ & $2.99(-06)$ & $0.00(+00)$ & $9.97(-01)$ & $0.0000(+00)$ & {\tt R\ 333-Y\ 111} & $4$ \\
$40.875$ & $.30333$ & $7.99(-03)$ & $5.76(+00)$ & $1.39(-03)$ & $1.5958(-02)$ & {\tt C\ 232-S\ 212} & $5$ \\
 \\
$40.876$ & $.30332$ & $7.70(-03)$ & $5.78(+00)$ & $1.33(-03)$ & $3.0744(-02)$ & {\tt C\ 234-S\ 212} & $6$ \\
\end{longtable}

\end{center}

%%%\end{document}

\newpage
\normalsize
Table IIa. Transitions 2l'3l"-1s2l,1s3l in [He] and 2p-1s, 3p-1s in [H] ions
\scriptsize
%TCIDATA{Version=5.00.0.2606}
%TCIDATA{LaTeXparent=0,0,arxiv.tex}

%%%\scriptsize

\begin{center}

\begin{longtable}{cccccccr}
%%% \caption{Table caption} \\

\hline
\hline
  WL, \AA & E, keV & A & G & K & QD & Transition & No \\
\hline
\endfirsthead

\hline
  WL, \AA & E, keV & A & G & K & QD & Transition & No \\
\hline
\endhead

\hline
\multicolumn{8}{r}{Continued on next page} \\
\hline
\endfoot

\hline \hline
\endlastfoot

\multicolumn{8}{l}{$  n= 3 \rightarrow n= 1;  Z= 6;$} \\
$28.465$ & $.43557$ & $2.17(-02)$ & $0.00(+00)$ & $1.00(+00)$ & $0.0000(+00)$ & {\tt R'234-Y\ 212} & $1$ \\
$28.466$ & $.43555$ & $2.17(-02)$ & $0.00(+00)$ & $1.00(+00)$ & $0.0000(+00)$ & {\tt R'232-Y\ 212} & $2$ \\
$29.288$ & $.42333$ & $3.47(-03)$ & $1.86(-03)$ & $4.72(-02)$ & $8.7803(-05)$ & {\tt A'331-S\ 313} & $3$ \\
$29.288$ & $.42333$ & $3.49(-03)$ & $2.20(-03)$ & $4.72(-02)$ & $3.1174(-04)$ & {\tt A'333-S\ 313} & $4$ \\
$29.289$ & $.42332$ & $3.52(-03)$ & $1.91(-03)$ & $4.79(-02)$ & $4.5781(-04)$ & {\tt A'335-S\ 313} & $5$ \\
 \\
$29.573$ & $.41925$ & $2.19(-03)$ & $4.17(-01)$ & $4.43(-03)$ & $5.5349(-03)$ & {\tt A'133-S\ 111} & $6$ \\
\end{longtable}

\end{center}

%%%\end{document}

\newpage
\normalsize
Table IIb. Transitions 1s2l'3l"-1s2.2l,1s2.3l in [Li] and 1s2p-1s2, 1s3p-1s2 in [He] ions
\scriptsize
%TCIDATA{Version=5.00.0.2606}
%TCIDATA{LaTeXparent=0,0,arxiv.tex}

%%%\scriptsize

\begin{center}

\begin{longtable}{cccccccr}
%%% \caption{Table caption} \\

\hline
\hline
  WL, \AA & E, keV & A & G & K & QD & Transition & No \\
\hline
\endfirsthead

\hline
  WL, \AA & E, keV & A & G & K & QD & Transition & No \\
\hline
\endhead

\hline
\multicolumn{8}{r}{Continued on next page} \\
\hline
\endfoot

\hline \hline
\endlastfoot

\multicolumn{8}{l}{$  n= 3 \rightarrow n= 1;  Z= 6;$} \\
$34.970$ & $.35454$ & $2.64(-02)$ & $0.00(+00)$ & $1.00(+00)$ & $0.0000(+00)$ & {\tt R'133-Y\ 111} & $1$ \\
$35.066$ & $.35357$ & $8.82(-07)$ & $0.00(+00)$ & $9.89(-01)$ & $0.0000(+00)$ & {\tt R'333-Y\ 111} & $2$ \\
$35.572$ & $.34855$ & $3.13(-04)$ & $9.07(-03)$ & $3.33(-03)$ & $1.2085(-04)$ & {\tt A'234-S\ 212} & $3$ \\
$35.572$ & $.34855$ & $3.14(-04)$ & $9.17(-03)$ & $3.34(-03)$ & $6.1295(-05)$ & {\tt A'232-S\ 212} & $4$ \\
$35.880$ & $.34555$ & $6.68(-04)$ & $2.43(-03)$ & $5.17(-02)$ & $2.5058(-04)$ & {\tt I'232-S\ 212} & $5$ \\
 \\
$35.881$ & $.34554$ & $6.76(-04)$ & $2.37(-03)$ & $5.18(-02)$ & $4.9136(-04)$ & {\tt I'234-S\ 212} & $6$ \\
\end{longtable}

\end{center}

%%%\end{document}

\normalsize

\end{document}